\documentclass[
 reprint,
 superscriptaddress,
 amsmath,amssymb,
 aps,
 pra
]{revtex4-2}

\usepackage{graphicx}
\usepackage{dcolumn}
\usepackage{bm}
\usepackage{tikz}
\usetikzlibrary{quantikz, decorations.pathreplacing}
\usepackage[colorlinks=true,linkcolor=blue,citecolor=blue,urlcolor=blue]{hyperref}
\usepackage{graphicx, amsmath, amssymb, braket, physics, mathtools}
\usepackage{mathrsfs}

\begin{document}

\preprint{APS/123-QED}

\title{A Path-Superposition Framework for Quantum Gate Teleportation}

\author{Santiago Ávila}
 \affiliation{Tecnologico de Monterrey, School of Engineering and Sciences, 14380, Mexico City, Mexico}
\author{Marco Enríquez}%
 \email{menriquezf@tec.mx}
\affiliation{Tecnologico de Monterrey, School of Engineering and Sciences, 01389, Santa Fe, Mexico 
}

\date{\today}

\begin{abstract}
Quantum gate teleportation enables distant parties to implement nonlocal quantum operations without physically transferring the participating qubits, making it a promising primitive for distributed quantum computing. We introduce a general framework for deterministic quantum gate teleportation based on path superposition, in which the target nonlocal operation is specified through the phase of a preshared maximally entangled resource and a suitable family of path-dependent local unitary operators. The framework establishes general design conditions that guarantee deterministic teleportation after measurement of the control qubits and the application of local correction operations. As representative realizations, we construct teleportation protocols for controlled-NOT (CNOT) and controlled-$Z$ (CZ) gates, demonstrating that different nonlocal operations can be implemented within the same protocol architecture through appropriate choices of the design parameters. We further outline a proof-of-concept photonic realization based on spatial-path and polarization degrees of freedom. The proposed framework identifies path superposition as a versatile resource for quantum gate teleportation and distributed quantum information processing.
\end{abstract}

\maketitle

\section{Introduction}
Quantum teleportation was first described by Bennett \textit{et al.}~\cite{bennett} as the transfer of an observer's unknown quantum state to another observer via one classical and one quantum channel over a physically significant distance. The process makes use of resources similar to the encoding and decoding principles proposed by Bennett and Wiesner~\cite{bennettAndWeisner} in dense coding. Quantum teleportation has made substantial progress since then, as it is now possible to teleport quantum gates, making it very attractive in the quantum computation field~\cite{EisertQGT, LiuCNOT, Main_2025}. Yet the feasibility of such schemes depends on the physical limits of current quantum hardware. 

The number of qubits in a quantum computer heavily impacts its ability to perform complex algorithms. Effective quantum algorithms that demonstrate quantum supremacy over classical algorithms, such as Shor’s algorithm and several modern proposals, tend to require many more qubits than currently possible on a single device ~\cite{vision_and_challenges,the_quantum_internet,Shor_1997,mohseni2024build}. However, qubit scaleup is mainly hindered by noise, which induces decoherence of quantum states~\cite{vision_and_challenges,the_quantum_internet,a_survey}. These effects, together with increased crosstalk and slower gate times, are the reasons why monolithic quantum chips with thousands of qubits pose issues in seven of the main current quantum computing technologies, including ion traps and superconductors ~\cite{the_path,vision_and_challenges,Bruzewicz_2019}. Quantum error correction becomes essential in such models, as each logical qubit must be mapped into several physical ones to counteract decoherence, an engineering challenge far greater than in classical systems even for tens of qubits~\cite{vision_and_challenges,Klimov_2024}.

Distributed quantum computing (DQC) offers an attractive alternative to monolithic chips, as it is scalable via task distribution among several processing modules~\cite{Cirac_1999,AbuGhanem2025}. To circumvent losses incurred during direct quantum state transfer between modules due to decoherence, quantum teleportation provides a loss-tolerant alternative~\cite{a_survey,Main_2025}. Extending this idea further, an even more powerful technique is quantum gate teleportation (QGT), which enables non-local two-qubit gates, allowing control and target qubits to be in different modules without the need to physically transfer the states~\cite{Main_2025}. 

The first QGT protocol was proposed by Eisert \textit{et al.}~\cite{EisertQGT}. In this scheme a CNOT gate was teleported, representing a significant breakthrough, as the combination of a CNOT gate with single qubit transformations suffices to achieve a global set of multi-qubit unitary transformations. However, the original protocol by Eisert \textit{et al.} relies on local CNOT operations, as do most QGT and teleportation-based schemes. Alternative formulations achieving universality through different gate sets have also been investigated, such as three-CZ-based schemes~\cite{Main_2025}. Among these, indefinite causal order (ICO) exemplifies how superposing orders can extend the conventional circuit model. By relaxing the conventional assumption of a fixed causal order, where each operation has a well-defined past and future~\cite{Oreshkov2012}, entangled control qubits can be used to implement causally nonseparable processes. This enables a coherent superposition of the black-box operations $A$ and $B$ in both orders, $AB$ and $BA$, as formalized in the quantum switch introduced by Chiribella \textit{et al.}~\cite{chiribella}. Liu \textit{et al.}~\cite{LiuCNOT} effectively utilized ICO relationships mediated by a switch to apply a non-local CNOT gate across a physically significant distance over an optical realization via photonic degrees of freedom (DOF).

A conceptually related yet physically distinct approach is path superposition, also called trajectory superposition, in which a particle may undergo one of two quantum channels, mediated by a coherent control ~\cite{kraus}. This resource has been used as a means of reducing noise in the transmission of quantum information~\cite{Rubino_2021,Chiribella_2019,Gisin_2005} and in a modified version of the conventional teleportation protocol, proving to have performance enhancements over previous protocols~\cite{kraus}. 

In this work, we introduce a general framework for quantum gate teleportation based on path superposition. The framework formulates gate teleportation as the problem of selecting the phase of a preshared maximally entangled resource together with a suitable family of path-dependent local unitary operations satisfying a set of general design conditions. To illustrate its capabilities, we construct deterministic teleportation protocols for a controlled-NOT (CNOT) gate and a controlled-$Z$ (CZ) gate between two spatially separated parties. Both the CNOT and CZ gates are elements of the Clifford group $\mathcal{C}^{(2)}$. While Clifford gates alone are not universal, their combination with any local gate from the third level of the Clifford hierarchy yields a universal gate set~\cite{Cui_2017}. Similar to previous QGT schemes, the path-dependent local unitary operations are mediated by a preshared pair, resulting in coherently controlled paths. After Hadamard-basis measurements of the control qubits and the corresponding deterministic local corrections, the protocol reproduces the action of the desired nonlocal gate up to a physically irrelevant global phase. The proposed CZ realization makes use of generalized phase gates, known for their experimental simplicity and robustness~\cite{Nakata_2014}. These diagonal phase operations can often be implemented virtually through software-defined phase offsets, making them attractive for quantum-information processing platforms~\cite{McKay_2017,Vezvaee_2025}. Finally, we outline a proof-of-concept photonic realization of the framework based on spatial-path and polarization degrees of freedom. More generally, the proposed formulation identifies path superposition as a flexible resource for quantum gate teleportation, providing an alternative to conventional schemes based on local entangling operations or indefinite causal order.

The organization of the paper is as follows. Section~\ref{GenFram} introduces the general framework for quantum gate teleportation via path superposition. Sections~\ref{cnotSec} and \ref{czSec} present its application to the teleportation of CNOT and controlled-$Z$ gates, respectively. A proof-of-concept photonic realization is discussed in Sec.~\ref{pcSec}, while Secs.~\ref{discSec} and \ref{concSec} present the discussion and conclusions.  

\section{General Framework for Quantum Gate Teleportation via Path Superposition}\label{GenFram}

\begin{figure}[ht]
\centering
\includegraphics[width=0.8\linewidth, trim={0cm 7cm 0cm 1cm}, clip]{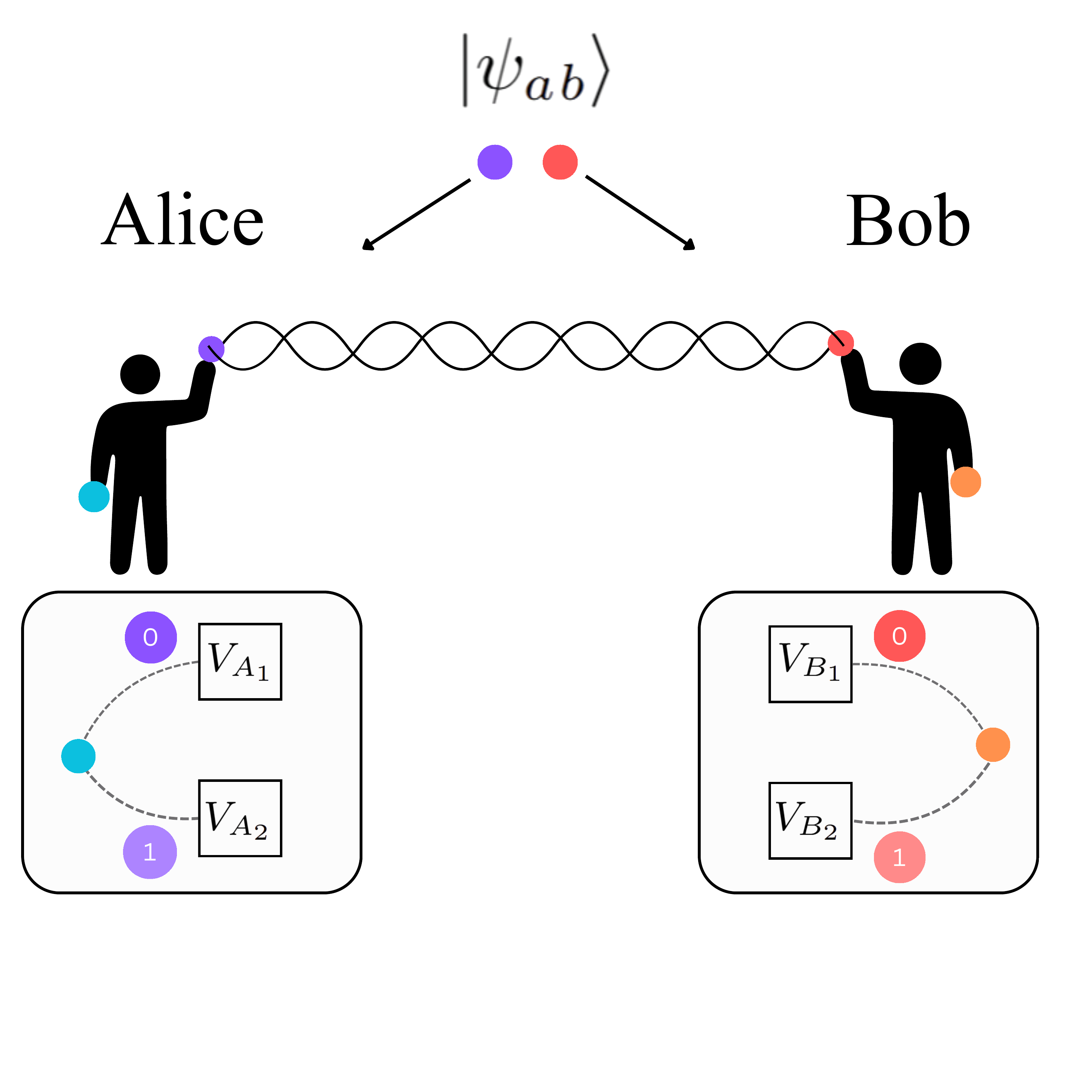}
\caption{Schematic diagram depicting distribution of qubits and the corresponding coherently controlled operations between Alice and Bob. Both parties possess one local qubit, $\ket{\psi_A}$ and $\ket{\psi_B}$ (blue and orange), and one entangled resource $\ket{\psi_a{_b}}$. The measurement of this control resource mediates the superposition of paths, causing $\ket{\psi_A}$ ($\ket{\psi_B}$) to undergo the transformation $V_{A_1}$ ($V_{B_1}$) or $V_{A_2}$ ($V_{B_2}$).}
\label{fig:AliceAndBob}
\end{figure}

We first introduce a general framework for teleporting nonlocal two-qubit gates by exploiting path superposition as a quantum resource. The protocol considers two distant observers, Alice and Bob, each holding an arbitrary input qubit together with one qubit of a preshared maximally entangled pair. The entangled resource coherently controls the application of two local unitary operations on each observer's qubit. This situation is depicted in Fig.~\ref{fig:AliceAndBob} .After Hadamard transformations on the control qubits and a projective measurement in the computational basis, the resulting state collapses into one of two conditional branches. Appropriate local corrections then map both branches onto the same target nonlocal operation.

Alice and Bob initially possess the arbitrary qubit states
\begin{align}\label{eq:AandBstates}
|\psi_A\rangle
&=
\sqrt{\alpha_A}|0\rangle+
\sqrt{1-\alpha_A}|1\rangle,
\nonumber\\
|\psi_B\rangle
&=
\sqrt{\alpha_B}|0\rangle+
\sqrt{1-\alpha_B}|1\rangle,
\end{align}
where $\alpha_A,\alpha_B\in[0,1]$. The coherent control is provided by the maximally entangled state
\begin{equation}
|\chi_{ab}(\eta)\rangle=
\frac{1}{\sqrt2}
\left(
|0_a0_b\rangle+
e^{i\eta}|1_a1_b\rangle
\right),
\end{equation}
where the parameter $\eta$ determines the relative phase between the two coherent paths. The complete initial state is therefore
\begin{equation}
|\Psi_0\rangle=
|\chi_{ab}(\eta)\rangle
|\psi_A\rangle
|\psi_B\rangle.
\end{equation}
Different values of $\eta$ may be employed depending on the gate to be teleported. In the applications presented in this work, the CNOT protocol uses $\eta=\pi/2$, whereas the controlled-$Z$ protocol employs $\eta=0$.

Each observer performs one of two local unitary operations conditioned on the corresponding control qubit,
\begin{align}\label{eq:operators}
O_A
&=
|0_a\rangle\langle0_a|
\otimes
V_{A_1}
+
|1_a\rangle\langle1_a|
\otimes
V_{A_2},
\\
O_B
&=
|0_b\rangle\langle0_b|
\otimes
V_{B_1}
+
|1_b\rangle\langle1_b|
\otimes
V_{B_2},
\end{align}
where the operators
$V_{A_1}$,
$V_{A_2}$,
$V_{B_1}$,
and
$V_{B_2}$
are selected according to the desired nonlocal gate.Applying these controlled operations yields
\begin{align}\label{eq:superpos}
\ket{\Psi}
= \frac{1}{\sqrt{2}}\Big[ &
  \ket{0_a}\ket{0_b}\!\otimes\!V_{A_1}\ket{\psi_A}\!\otimes\!V_{B_1}\ket{\psi_B} \nonumber\\
  +e^{i\eta}\,&
  \ket{1_a}\ket{1_b}\!\otimes\!V_{A_2}\ket{\psi_A}\!\otimes\!V_{B_2}\ket{\psi_B}
\Big].
\end{align}

The entangled control therefore generates a coherent superposition of two independent pairs of local operations. Both observers subsequently perform Hadamard transformations on their control qubits, converting the coherent path information into interference between the two branches. Measuring the control qubits in the computational basis projects the system onto one of two conditional branches,
\begin{equation}\label{eq:branchpp}
|\psi_{\pm\pm}\rangle
=
\frac{1}{\sqrt{{\cal N}_{\pm\pm}}}
\left(
V_{A_1}\otimes V_{B_1}
+
e^{i\eta}
V_{A_2}\otimes V_{B_2}
\right)
|\psi_A\rangle
|\psi_B\rangle,
\end{equation}
\begin{equation}\label{eq:branchpm}
|\psi_{\pm\mp}\rangle
=
\frac{1}{\sqrt{{\cal N}_{\pm\mp}}}
\left(
V_{A_1}\otimes V_{B_1}
-
e^{i\eta}
V_{A_2}\otimes V_{B_2}
\right)
|\psi_A\rangle
|\psi_B\rangle.
\end{equation}

The construction of a gate teleportation protocol within this framework therefore reduces to selecting the phase $\eta$ together with the local unitary operators
$V_{A_1}$,
$V_{A_2}$,
$V_{B_1}$,
and
$V_{B_2}$
such that the following design conditions are satisfied:

\begin{enumerate}
\item The conditional states are properly normalized.

\item Both measurement branches occur with equal probability.

\item Suitable local corrections transform both conditional branches into the same target nonlocal gate, up to an overall physically irrelevant global phase.
\end{enumerate}

The first two design conditions can be analyzed by introducing the operator

\begin{equation}
W=
V_{A_1}^{\dagger}V_{A_2}
\otimes
V_{B_1}^{\dagger}V_{B_2},
\end{equation}

and its expectation value

\begin{equation}
z=
\langle\psi_A|
\langle\psi_B|
W
|\psi_A\rangle
|\psi_B\rangle.
\end{equation}

\begin{figure*}[htbp]
\centering
\includegraphics[width=0.8\linewidth, trim={0cm 1cm 0cm 0cm}, clip]{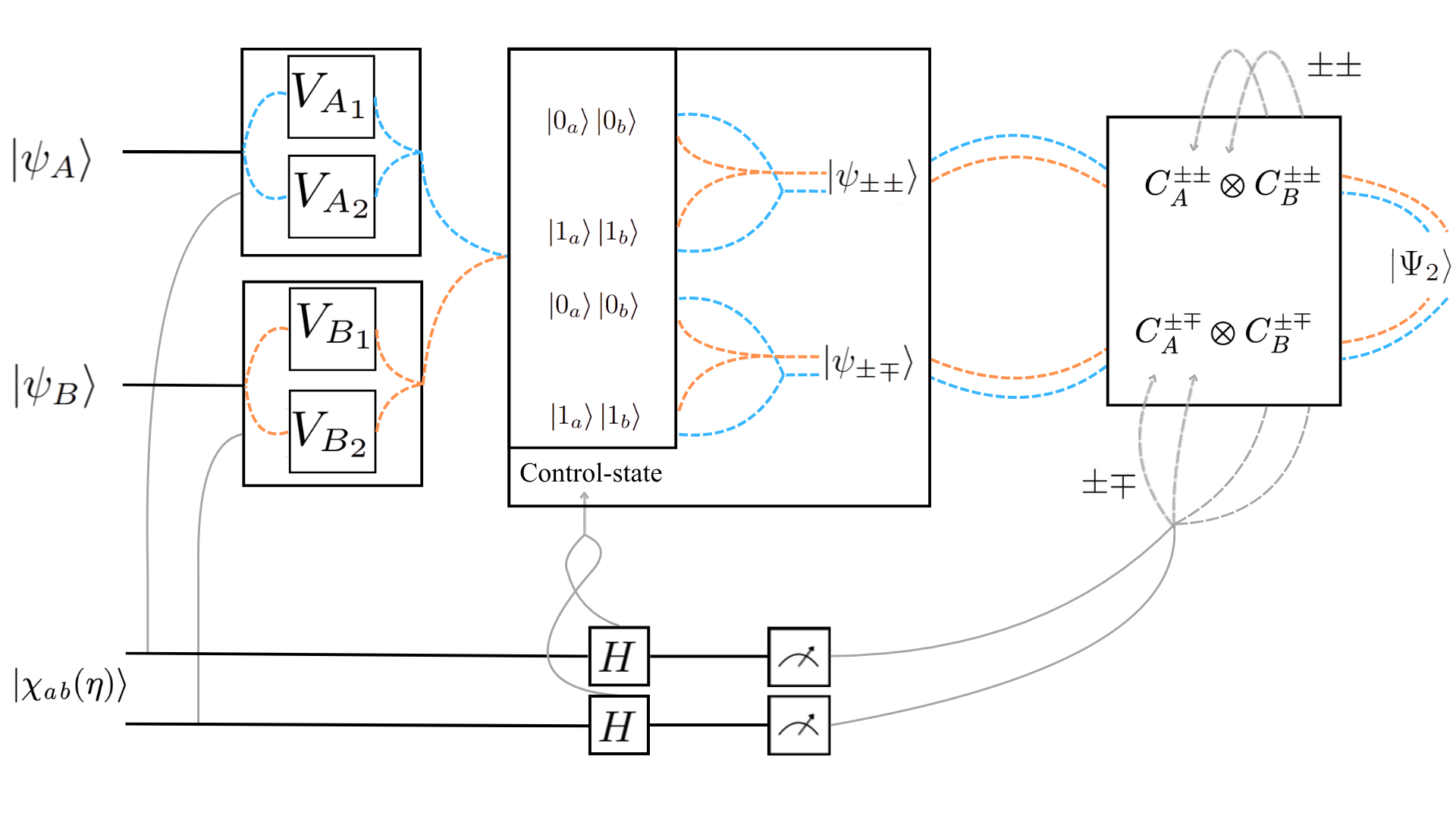}
\caption{Schematic representation of the full teleportation protocol. Coherent control is illustrated by showing the two possible orderings that $\ket{\psi_A}$ and ($\ket{\psi_B}$) undergo, mediated by the entangled control pair (gray wires). This generates a superposition of paths, depicted by the dashed blue and orange wires. A Hadamard gate is then applied to each control qubit, transforming the initial control superposition of $\ket{00}$ and $\ket{11}$ to $\ket{00}$, $\ket{11}$, $\ket{10}$, and $\ket{01}$, represented by the gray wires mediating the superpositions in the control-state. A measurement of the control qubit collapses the system onto one of the two Alice and Bob subsystems, allowing them to perform the appropriate correction. The final set of gray arrows show the two possible outcomes ($\pm\pm$ or $\pm\mp$) and the correction paths associated to them, with the $C^{\pm\pm}$ and  $C^{\pm\mp}$ operators representing the respective set of corrections for each protocol. Finally, both sets of corrections map the system to the same final state $\ket{\Psi_2}$.}
\label{fig:fullProcess}
\end{figure*}

Thus, the normalization factors become

\begin{align}
{\cal N}_{\pm\pm}
&=
2+
2\,\mathrm{Re}
\left(
e^{i\eta}z
\right),
\\
{\cal N}_{\pm\mp}
&=
2-
2\,\mathrm{Re}
\left(
e^{i\eta}z
\right),
\end{align}
from which the probabilities associated with each measurement branch are

\begin{align}
p_{\pm\pm}
&=
\frac{{\cal N}_{\pm\pm}}{4}
=
\frac12
\left[
1+
\mathrm{Re}
\left(
e^{i\eta}z
\right)
\right],
\\
p_{\pm\mp}
&=
\frac{{\cal N}_{\pm\mp}}{4}
=
\frac12
\left[
1-
\mathrm{Re}
\left(
e^{i\eta}z
\right)
\right].
\end{align}
Therefore, equally probable measurement branches are obtained whenever
\begin{equation}\label{eq:eqprobCond}
\mathrm{Re}
\left(
e^{i\eta}z
\right)=0.
\end{equation}
This provides a state-independent criterion for selecting admissible families of local unitary operators. Once this condition is satisfied, the remaining design condition depends only on the target nonlocal gate and determines the corresponding local operators together with the required correction operations. In Fig.\ref{fig:fullProcess} we depict the whole process described above.

The framework naturally defines a systematic procedure for constructing path-superposition-assisted gate teleportation protocols. Given a target nonlocal gate, the problem reduces to selecting the phase of the entangled resource together with a family of path-dependent local unitary operators satisfying the design conditions established above. Once these conditions are fulfilled, the corresponding local correction operations are determined by the conditional output branches. The following sections illustrate this procedure through the construction of deterministic teleportation protocols for the CNOT and controlled-$Z$ gates.
\section{Application I: Teleportation of the Controlled-NOT Gate}\label{cnotSec}

We now apply the framework introduced in Sec.~\ref{GenFram} to the construction of a deterministic teleportation protocol for a controlled-NOT (CNOT) gate. Following the general design procedure, we first specify the target operation, then select the phase of the entangled resource and a family of local unitary operators satisfying the design conditions. Finally, we determine the local correction operations required to recover the target nonlocal gate from each conditional branch.

The target operation is the action of a CNOT gate between Alice's and Bob's input qubits, with Alice's qubit acting as the control and Bob's qubit as the target. Thus, for the input state
$|\psi_{AB}\rangle=|\psi_A\rangle|\psi_B\rangle$, the desired transformation is
\begin{equation}
\left(
|0\rangle\langle0|\otimes I
+
|1\rangle\langle1|\otimes X
\right)
|\psi_{AB}\rangle .
\end{equation}
Using the definitions of $|\psi_A\rangle$ and $|\psi_B\rangle$ given in equation (\ref{eq:AandBstates}):
\begin{eqnarray}\label{eq:cnot}
&&|\Psi_{\mathrm{CNOT}}\rangle\nonumber\\[6pt]
 & &=\sqrt{\alpha_A\alpha_B}\ket{00}
   +\sqrt{\alpha_A(1-\alpha_B)}\ket{01}\nonumber \\[6pt]
   \quad& &+\sqrt{(1-\alpha_A)(1-\alpha_B)}\ket{10}
   +\sqrt{(1-\alpha_A)\alpha_B}\ket{11}. \quad\quad
\end{eqnarray}
For the CNOT protocol we choose the phase of the entangled resource to be $\eta=\pi/2$

such that the shared Bell state becomes
\begin{equation}\label{eq:entangled}
    \ket{\psi_a{_b}}=\frac{1}{\sqrt{2}}\big(\ket{0_a0_b}+i\ket{1_a1_b}\big).
\end{equation}
According to the framework developed in Sec.~\ref{GenFram}, the remaining task is to determine a family of local unitary operators satisfying the three design conditions. In particular, the selected local unitary operators must satisfy the equal-probability condition $\mathrm{Re}\!\left(e^{i\eta}z\right)=0$,
which, for $\eta=\pi/2$, reduces to $\mathrm{Im}(z)=0.$ One admissible family of local unitary operators satisfying these conditions is
\begin{equation}\label{eq:matrices}
\begin{array}{cc}
V_{A_1} = \begin{pmatrix} 0 & 1 \\[2pt] 1 & 0 \end{pmatrix},
&
V_{A_2} = \begin{pmatrix} 0 & -i \\[2pt] i & 0 \end{pmatrix},
\\[1.9em]
V_{B_1} = \dfrac{1}{\sqrt{2}}\begin{pmatrix} ie^{i\sigma} & e^{i\sigma} \\[2pt] e^{i\tau} & ie^{i\tau} \end{pmatrix},
&
V_{B_2} = \dfrac{1}{\sqrt{2}}\begin{pmatrix} ie^{i\sigma} & -e^{i\sigma} \\[2pt] -e^{i\tau} & ie^{i\tau} \end{pmatrix},
\\[1em]
\end{array}
\end{equation}
where $\sigma$ and $\tau$ are arbitrary real parameters. Notice that
$V_{A_1}=X$ and $V_{A_2}=Y$, while $V_{B_1}$ and $V_{B_2}$ remain unitary for any choice of $\sigma$ and $\tau$. The above operators constitute one representative realization of the proposed framework. Other families of local unitary operators satisfying the same design conditions may also reproduce the target CNOT operation after the corresponding local corrections. Furthermore, these operators satisfy the equal-probability condition derived (\ref{eq:eqprobCond}), ensuring that both measurement branches occur with probability $1/2$. Substituting the operators above into the conditional branches (\ref{eq:branchpp}) and (\ref{eq:branchpm}), the normalization factors become
$
{\cal N}_{\pm\pm}={\cal N}_{\pm\mp}=2,
$
reflecting that both conditional branches occur with equal probability. The resulting conditional branches reduce to
\begin{equation}\label{eq:branch_plus_expanded}
\begin{split}
\ket{\psi_\pm{_\pm}}=\big[&ie^{i\sigma}\sqrt{(1-\alpha_A)\alpha_B}\ket{00}
    \\
    +&ie^{i\tau}\sqrt{(1-\alpha_A)(1-\alpha_B)}\ket{01}
    \\
    +&e^{i\sigma}\sqrt{\alpha_A(1-\alpha_B)}\ket{10}
    \\
    +&e^{i\tau}\sqrt{\alpha_A\alpha_B} \ket{11}\big],
    \\
\end{split}
\end{equation}
and
\begin{equation}\label{eq:branch_min_expanded}
\begin{split}
\ket{\psi_\pm{_\mp}}=\big[&e^{i\sigma}\sqrt{(1-\alpha_A)(1-\alpha_B)}\ket{00}
    \\
    +&e^{i\tau}\sqrt{(1-\alpha_A)\alpha_B}\ket{01}
    \\
    +&ie^{i\sigma}\sqrt{\alpha_A\alpha_B}\ket{10}
    \\
    +&ie^{i\tau}\sqrt{\alpha_A(1-\alpha_B)}\ket{11}\big]
    \\
\end{split}
\end{equation}
Neither conditional branch coincides directly with the target state (\ref{eq:cnot}). Nevertheless, both branches differ from the desired transformation only by local permutations of the computational basis and local phase factors. Consequently, deterministic teleportation is completed by applying local correction operations conditioned on the measurement outcome. For the branch (\ref{eq:branch_plus_expanded}), the required corrections are

\begin{equation}
\mathrm{Alice}:
\;
R_z\!\left(-\frac{\pi}{2}\right)X,
\qquad
\mathrm{Bob}:
\;
R_z(\tau-\sigma)X,
\end{equation}
whereas for the branch (\ref{eq:branch_min_expanded}) they become
\begin{equation}
\mathrm{Alice}:
\;
R_z\!\left(\frac{\pi}{2}\right)X,
\qquad
\mathrm{Bob}:
\;
R_z(\sigma-\tau).
\end{equation}
Applying the corresponding local corrections transforms both conditional branches into the same final state
\begin{equation}
\begin{split}
\ket{\Psi_2}=e^{i(\frac{\pi}{4}+\frac{\sigma+\tau}{2})}\big[&\sqrt{\alpha_A\alpha_B}\ket{00}
    \\
    +&\sqrt{\alpha_A(1-\alpha_B)}\ket{01}
    \\
    +&\sqrt{(1-\alpha_A)(1-\alpha_B)}\ket{10}
    \\
    +&\sqrt{(1-\alpha_A)\alpha_B}\ket{11}\big],
    \\
\end{split}
\end{equation}
which differs from the target state only by an overall physically irrelevant global phase. Therefore, the selected entangled resource and local unitary operators satisfy the three design conditions established in Sec.~\ref{GenFram}, demonstrating that the proposed path-superposition protocol deterministically teleports a nonlocal controlled-NOT gate.

\section{Application II: Teleportation of the Controlled-$Z$ Gate}\label{czSec}
The same design procedure can be applied to construct a deterministic teleportation protocol for a controlled-$Z$ gate. Although the target operation differs from the CNOT gate, the overall protocol architecture remains unchanged. Only the phase of the entangled resource and the corresponding family of local unitary operators must be modified to satisfy the design conditions established in Sec.~\ref{GenFram}. The target operation is
\begin{equation}
\left(
|0\rangle\langle0|\otimes I
+
|1\rangle\langle1|\otimes Z
\right)
|\psi_{AB}\rangle ,
\end{equation}
which transforms the input state into
\begin{eqnarray}\label{eq:cz}
&&|\Psi_{\mathrm{CZ}}\rangle
=\sqrt{\alpha_A\alpha_B}\ket{00}
   +\sqrt{\alpha_A(1-\alpha_B)}\ket{01}\nonumber\\[6pt]
   \quad& &+\sqrt{(1-\alpha_A)\alpha_B}\ket{10}
   -\sqrt{(1-\alpha_A)(1-\alpha_B)}\ket{11}. \quad\quad
\end{eqnarray}
Unlike the CNOT protocol, where the desired transformation exchanges computational basis states, the CZ gate only introduces a conditional phase. Consequently, it is sufficient to restrict the local operators to diagonal unitary matrices, preserving the computational basis while modifying only the relative phases of the computational basis amplitudes. For this realization we choose $\eta=0$, such that the entangled resource becomes
\begin{equation}
\vert \varphi_{ab}\rangle=\frac{1}{\sqrt{2}}\left(|0_a0_b\rangle+|1_a1_b\rangle\right).
\end{equation}
With this choice, the equal-probability condition derived in Sec.~\ref{GenFram} reduces to $\mathrm{Re}(z)=0$. To satisfy the remaining design conditions, we consider the family of diagonal local operators
\begin{equation}\label{eq:matrices_CZ}
\begin{array}{cc}
V_{A_1} = \begin{pmatrix} \psi_1 & 0 \\[2pt] 0 & 1 \end{pmatrix},
&
V_{A_2} = \begin{pmatrix} \psi_2 & 0 \\[2pt] 0 & 1 \end{pmatrix},
\\[1.5em]
V_{B_1} = \begin{pmatrix} a & 0 \\[2pt] 0 & b \end{pmatrix},
&
V_{B_2} = \begin{pmatrix} c & 0 \\[2pt] 0 & d \end{pmatrix},
\end{array}
\end{equation}
where $a$, $b$, $c$, $d$, $\psi_1$, and $\psi_2$ are complex phases of unit modulus. The normalization condition introduced in Sec.~\ref{GenFram} imposes restrictions on the free parameters of the local operators. Substituting this family of operators into the normalization factors yields
\begin{equation}\label{eq:normcond}
a\overline{c}+\overline{a}c=b\overline{d}+\overline{b}d=0,\quad \psi_1{\overline \psi_2}={\overline \psi_1}\psi_2.
\end{equation}
where the overline denotes complex conjugation. Since all parameters have unit modulus, these relations constrain only their relative phases. We next analyze the conditional branch $|\psi_{\pm\mp}\rangle$. Substituting the operators into the general expression derived in Sec.~\ref{GenFram} gives
\begin{equation}\label{eq:branch_min_expanded_CZ}
\begin{split}
\ket{\psi_\pm{_\mp}}=\frac1{\sqrt 2}& \big[(a\psi_1 - c\psi_2)\sqrt{\alpha_A\alpha_B}\ket{00}
    \\
    +&(b\psi_1 - d\psi_2)\sqrt{\alpha_A(1-\alpha_B)}\ket{01}
    \\
    +&(a - c)\sqrt{(1-\alpha_A)\alpha_B}\ket{10}
    \\
    +&(b - d)\sqrt{(1-\alpha_A)(1-\alpha_B)}\ket{11}\big].
    \\
\end{split}
\end{equation}
To reproduce the target state $|\Psi_{\mathrm{CZ}}\rangle$, the four coefficients above must coincide with those of Eq.~(\ref{eq:cz}), up to an overall global phase. Since the CZ gate only changes the relative phase of the state $|11\rangle$, the first three amplitudes must be equal, while the fourth must acquire the opposite sign. This requirement leads to
\begin{equation}\label{eq:condCZ}
    a\psi_1 - c\psi_2=b\psi_1 - d\psi_2=d-b=a-c.
\end{equation}
Combining the normalization constraints with the conditions imposed by the target operation yields one admissible family of solutions,
\begin{equation}
a=d=1,
\qquad
b=c=-i,
\qquad
\psi_1=-i,
\qquad
\psi_2=i.
\end{equation}
As in the CNOT case, this choice is not unique but serves as a representative realization satisfying the design conditions established in Sec.~\ref{GenFram}.
Substituting these values into the conditional branch gives
\begin{equation}
    \begin{split}
\ket{\psi_\pm{_\mp}}=-e^{i\pi/4}& \big[ \sqrt{\alpha_A\alpha_B}\ket{00}
    \\
    +&\sqrt{\alpha_A(1-\alpha_B)}\ket{01}
    \\
    -&\sqrt{(1-\alpha_A)\alpha_B}\ket{10}
    \\
    +&\sqrt{(1-\alpha_A)(1-\alpha_B)}\ket{11}\big],
    \\
\end{split}
\end{equation}
which differs from the target state only by a local $Z$ operation on Alice's qubit together with an overall global phase. The second conditional branch is obtained by substituting the same family of operators into
$|\psi_{\pm\pm}\rangle$. This yields
\begin{equation}
    \begin{split}
\ket{\psi_\pm{_\pm}}=e^{-i\pi/4}& \big[ \sqrt{\alpha_A\alpha_B}\ket{00}
    \\
    -&\sqrt{\alpha_A(1-\alpha_B)}\ket{01}
    \\
    +&\sqrt{(1-\alpha_A)\alpha_B}\ket{10}
    \\
    +&\sqrt{(1-\alpha_A)(1-\alpha_B)}\ket{11},
    \\
\end{split}
\end{equation}
which differs from the target state only by a local Pauli-$Z$ operation acting on Bob's qubit and an overall physically irrelevant global phase. After applying the corresponding local correction, both conditional branches are transformed into a state which differs to (\ref{eq:cz}) up to a global phase.

The selected entangled resource and local unitary operators therefore satisfy the three design conditions established in Sec.~\ref{GenFram}. Consequently, after communicating the measurement outcome through two classical bits and applying the corresponding local correction, the protocol deterministically reproduces the action of a nonlocal controlled-$Z$ gate.

\section{Proof-of-Concept Photonic Realization}\label{pcSec}

\begin{figure*}[t]
\centering
\includegraphics[
    width=\textwidth,
    trim={0cm 7cm 0cm 1cm},
    clip
]{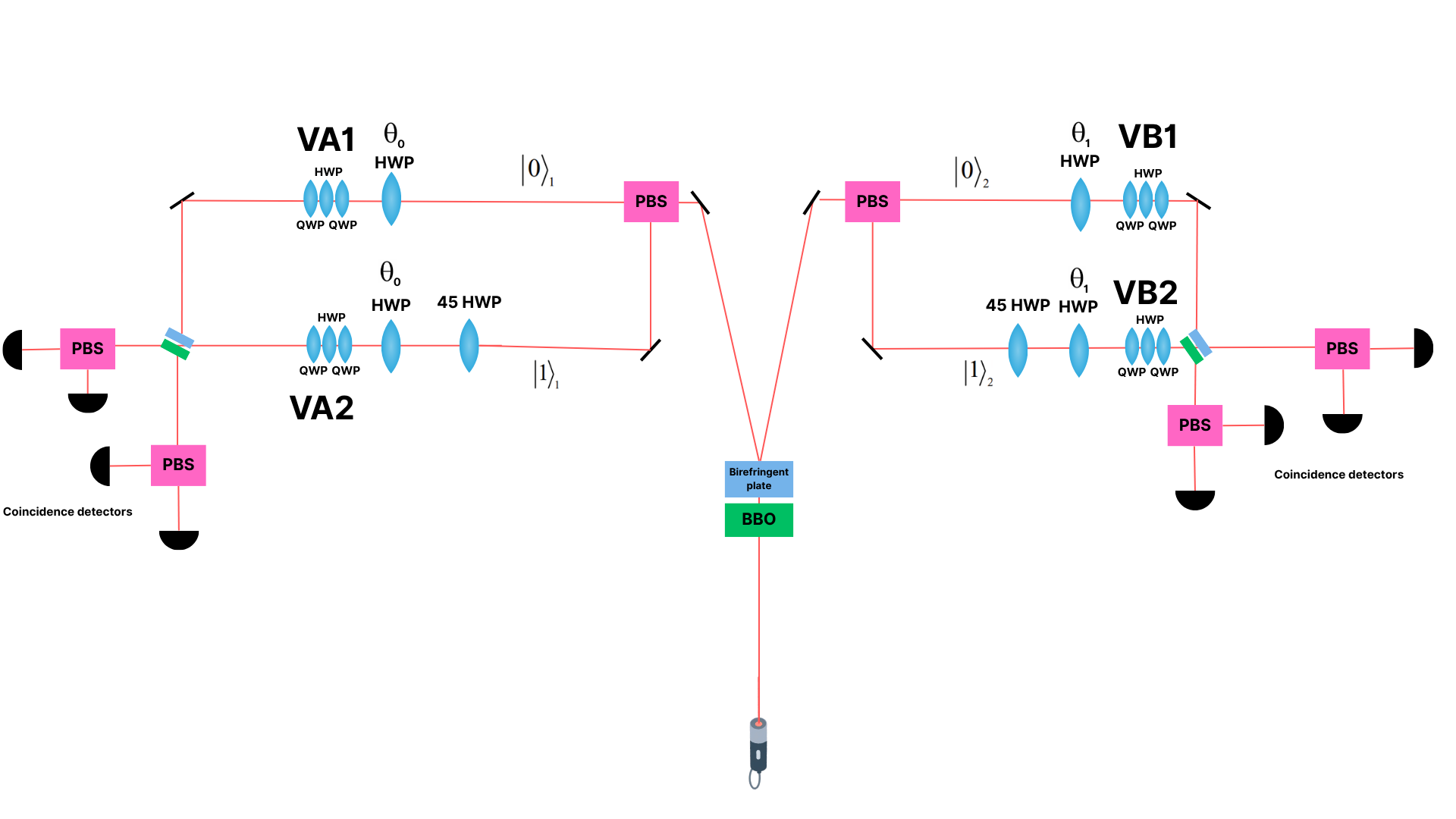}
\caption{Proof-of-concept photonic realization of the proposed framework. A BBO-based SPDC source generates a polarization-entangled photon pair \cite{kwiat}, PBSs map polarization into spatial path modes, and \(45^\circ\) HWPs transfer the entanglement to the path DOF \cite{LiuCNOT}. A HWP at angle ($\theta$) then prepares each polarization qubit in an arbitrary linear-polarization state. Additional HWP and QWP sequences implement path-dependent polarization operations that can perform any $SU(2)$ unitary transformation \cite{SIMON1990165}, therefore able to apply the V set gates. Recombination of the two spatial paths using a 50/50 beam splitter results in Hadamard like operations \cite{Procopio_2015}, replicating equation (\ref{eq:superpos}). The resulting coincidence branch determines the required local correction operations.}
\label{fig:opticalFigure}
\end{figure*}

The framework introduced in Sec.~\ref{GenFram} admits a natural proof-of-concept photonic realization based on the spatial-path and polarization degrees of freedom of single photons. In this implementation, the path degree of freedom encodes the coherent control provided by the entangled resource, while the polarization represents the computational qubit. The path-dependent local unitary operations are implemented independently on each interferometric branch, and the final interference of the paths, followed by projective measurements and local feed-forward corrections, reproduces the protocol architecture described in Sec.~\ref{GenFram}. Different choices of the path-dependent optical elements therefore realize different representative protocols within the same experimental configuration.
For the proof-of-concept photonic realization we consider a photonic system since photonic degrees of freedom (DOF) have been widely used to implement teleportation protocols and related ICO schemes~\cite{Procopio_2015,Rubino2022experimental,Rubino_2021,LiuCNOT}. The proposed setup follows an interferometric path-encoding approach, analogous to Mach–Zehnder-type architectures \cite{Procopio_2015,Rubino_2021}, in which coherent spatial modes represent path alternatives while polarization encodes the target qubits. For the CNOT and CZ teleportation protocols described above, the required qubits are encoded in distinct photonic degrees of freedom. These DOFs correspond to physical properties of the photon, such as polarization and spatial path, each associated with an independent Hilbert space that can be used to store quantum information. This allows multiple qubits to be encoded in different DOFs of the same photon \cite{dof}. Optical elements are then used to manipulate these encoded DOFs, reproducing the state transformations required by the abstract protocols. The aim of this section is to provide a proof of concept of the path-dependent transformations by showing this equivalence at the level of the encoded DOFs. The coincidence outcomes identify the state’s branch; yet nondestructive photonic-qubit detection methods that preserve the polarization DOF would be necessary in order to apply the local corrections and determine the final CNOT and CZ equivalent states \cite{Niemietz2021}. Possible optical elements implementing these transformations are shown in Figure~\ref{fig:opticalFigure}.

The proposed realization employs the polarization states
$|H\rangle$ and $|V\rangle$
to encode the computational basis of the input qubits, whereas the two spatial paths represent the coherent control associated with the entangled resource. Beam splitters generate and recombine the path superposition, while wave plates placed in each arm implement the local unitary operators required by the protocol.
The photonic realization begins with the generation of a maximally entangled photon pair in the polarization degree of freedom, where one photon is then assigned to each observer. $\ket{H}$ and $\ket{V}$ are denoted as horizontal and vertical polarization states. Such a polarization-entangled pair may be produced through spontaneous parametric down-conversion (SPDC) \cite{kwiat}:

\begin{equation}\label{eq:SPDC}
\ket{\Phi_0}=\frac{1}{\sqrt{2}}\big(\ket{H_a}\ket{{H_b}} + i\ket{V_a}\ket{{V_b}}\big)
\end{equation}

State~(\ref{eq:SPDC}) is equivalent to the entangled pair appearing in the abstract protocol, now encoded in the polarization Hilbert space, corresponding to a single qubit each. Note that the phase factor may be changed for each protocol by tilting BBO crystal seen in Figure \ref{fig:opticalFigure} \cite{kwiat}. In order to introduce a spatial DOF to each photon and therefore two new qubits, the observers must send their photon through a polarization-dependent path separation process in order to coherently route horizontal and vertical polarization into distinct spatial paths, introducing states $\ket{0}$ and $\ket{1}$ respectively in the spatial Hilbert space (\ref{eq:SPDC_and_spatial}).

\begin{equation}\label{eq:SPDC_and_spatial}
\ket{\Phi_1}=\frac{1}{\sqrt{2}}\big(\ket{H_a}\ket{0_a}\ket{{H_b}}\ket{0_b} + i\ket{V_a}\ket{1_a}\ket{{V_b}}\ket{1_b}\big)
\end{equation}

Entanglement in state (\ref{eq:SPDC_and_spatial}) is transferred to the spatial qubits by rotating the vertical polarization qubits $\ket{V}$ in the labeled paths $\ket{1}$ in order to make all polarization horizontal $\ket{H}$, resulting in state (\ref{eq:factoredPolarization}).

\begin{equation}\label{eq:factoredPolarization}
\ket{\Phi_2}=\frac{1}{\sqrt{2}}\big(\ket{0_a}\ket{0_b} + i\ket{1_a}\ket{1_b}\big) \otimes\big(\ket{{H_a}}\ket{{H_b}}\big)
\end{equation}

Once this factorized state has been obtained, arbitrary linear-polarization states may be prepared locally for each photon. The resulting state is shown in (\ref{eq:entangledAandArbitrary}) and consists of a maximally entangled pair encoded in the spatial DOF, together with two independent polarization qubits. This reproduces the initial resource structure required by the CNOT and CZ teleportation protocols, with the spatial DOF encoding the control qubits and the polarization DOF encoding Alice’s and Bob’s arbitrary target qubits.

\begin{equation}\label{eq:entangledAandArbitrary}
\ket{\Phi_3}=\frac{1}{\sqrt{2}}\big(\ket{0_a}\ket{0_b} + i\ket{1_a}\ket{1_b}\big) \ket{{\psi_A}}\ket{{\psi_B}}
\end{equation}

Gates are then applied to the polarization qubits by simply placing the optical elements on the corresponding spatial path, replicating the action of the operators found in (\ref{eq:operators}), resulting in state (\ref{eq:superposOPTICAL}) and an effective path superposition.   

\begin{align}\label{eq:superposOPTICAL}
\ket{\Phi_4}
= \frac{1}{\sqrt{2}}\Big[ &
  \ket{0_a}\ket{0_b}\!\otimes\!V_{A_1}\ket{\psi_A}\!\otimes\!V_{B{}_1}\ket{\psi_B} \nonumber\\
  +i\,&
  \ket{1_a}\ket{1_b}\!\otimes\!V_{A{}_2}\ket{\psi_A}\!\otimes\!V_{B{}_2}\ket{\psi_B}
\Big].
\end{align}
Finally, the spatial modes are coherently recombined locally by each observer, implementing a Hadamard-type basis change on the path qubits up to fixed optical phases. The resulting coincidence pattern identifies the joint path-measurement branch ($\pm\pm \text{or} \pm\mp$) and therefore determines the local corrections associated with the corresponding polarization output.
Within this implementation, the CNOT and controlled-$Z$ teleportation protocols correspond to different configurations of the local optical elements acting on each interferometric path. Consequently, the same experimental architecture can implement different teleported gates by modifying only the path-dependent unitary transformations and the phase of the entangled resource, without changing the overall structure of the protocol.
Although presented as a proof-of-concept realization, this implementation illustrates that the proposed framework is compatible with experimentally accessible linear-optical technologies. More generally, it highlights the modular nature of the protocol, in which the experimental architecture remains unchanged while different nonlocal gates are realized through appropriate choices of the path-dependent local operations.

\section{Discussion}\label{discSec}

An important conceptual aspect of the proposed framework is that it reformulates gate teleportation as a systematic design procedure. Rather than constructing an independent protocol for each nonlocal operation, the framework identifies the entangled-resource phase and the path-dependent local unitary operators as the fundamental design variables. The CNOT and controlled-$Z$ protocols demonstrate how different choices of these variables generate different teleported gates while preserving the same underlying protocol architecture.

From this perspective, the proposed framework complements conventional quantum gate teleportation schemes. Standard approaches typically rely on a prescribed sequence of local entangling operations, Bell-state measurements, and classical communication to implement a specific nonlocal gate. In contrast, the present approach exploits the coherent superposition of path-dependent local evolutions, where interference between the two branches, together with deterministic local corrections, gives rise to the desired nonlocal operation. Path superposition therefore constitutes an alternative physical resource for implementing distributed quantum gates.

Another noteworthy feature of the framework is the non-uniqueness of the local unitary operators. The operator families presented for the CNOT and controlled-$Z$ gates are representative realizations satisfying the design conditions introduced in Sec.~\ref{GenFram} rather than unique solutions. Alternative families of local operators may satisfy the same conditions and reproduce the target nonlocal gate after appropriate local corrections. This flexibility may facilitate the adaptation of the framework to different experimental platforms and implementation constraints while preserving the overall protocol architecture.

The proof-of-concept photonic realization further illustrates that the proposed framework can be implemented using existing optical technologies. By encoding the coherent control in spatial paths and the computational information in polarization, the required path-dependent local operations can be naturally implemented within a linear-optical architecture. Although presented only as a conceptual realization, this proposal demonstrates that the framework is compatible with experimentally accessible photonic techniques.

The present work is restricted to deterministic teleportation of two representative gates under ideal conditions. The framework does not yet provide a complete characterization of all admissible families of local unitary operators, nor does it address the effects of decoherence, imperfect measurements, or noisy communication channels. Extending the framework to these more realistic scenarios constitutes an interesting direction for future research.

Taken together, these results suggest that path superposition should be regarded not merely as a resource for implementing isolated teleportation protocols but as a general design principle for distributed quantum gate teleportation. This perspective naturally raises the broader question of identifying which classes of nonlocal gates admit realizations within the proposed framework and how such realizations can be systematically constructed.

The proposed framework should therefore be regarded as a systematic methodology for constructing path-superposition-assisted gate teleportation protocols rather than a construction tailored to a specific gate.

\section{Conclusions}\label{concSec}

We have introduced a general framework for quantum gate teleportation based on path superposition. The framework identifies the phase of a preshared maximally entangled resource and a family of path-dependent local unitary operations as the key design parameters for constructing deterministic teleportation protocols. The proposed design conditions provide a unified procedure for implementing nonlocal quantum gates through coherent path superposition and local correction operations.

As representative realizations of the framework, we constructed deterministic teleportation protocols for the CNOT and controlled-$Z$ gates and outlined a proof-of-concept photonic realization. These examples demonstrate the flexibility of the framework and show how different nonlocal operations can be realized within the same protocol architecture through suitable choices of the entangled resource and local operators.

The proposed framework opens the possibility of constructing teleportation protocols for additional classes of nonlocal gates and extending the formalism to noisy channels, higher-dimensional systems, and multipartite quantum networks, further establishing path superposition as a versatile resource for distributed quantum information processing.
\begin{acknowledgments}
Authors are grateful for the support to publish this article provided by the School of Engineering and Science, Tecnologico de Monterrey 
\end{acknowledgments}


\begin{thebibliography}{31}

\bibitem{bennett}
C.~H. Bennett, G. Brassard, C. Cr{\'e}peau, R. Jozsa, A. Peres, and W.~K. Wootters,
``Teleporting an unknown quantum state via dual classical and Einstein-Podolsky-Rosen channels,''
Phys. Rev. Lett. \textbf{70}, 1895 (1993).

\bibitem{bennettAndWeisner}
C.~H. Bennett and S.~J. Wiesner,
``Communication via one- and two-particle operators on Einstein-Podolsky-Rosen states,''
Phys. Rev. Lett. \textbf{69}, 2881--2884 (1992).

\bibitem{EisertQGT}
J. Eisert, K. Jacobs, P. Papadopoulos, and M.~B. Plenio,
``Optimal local implementation of nonlocal quantum gates,''
Phys. Rev. A \textbf{62}, 052317 (2000).

\bibitem{LiuCNOT}
W.-Q. Liu and H.-R. Wei,
``Quantum gate teleportation with the superposition of causal order,''
Phys. Rev. Appl. \textbf{23}, 014064 (2025).

\bibitem{Main_2025}
D. Main, P. Drmota, D.~P. Nadlinger, E.~M. Ainley, A. Agrawal, B.~C. Nichol, R. Srinivas, G. Araneda, and D.~M. Lucas,
``Distributed quantum computing across an optical network link,''
Nature \textbf{638}, 383--388 (2025).

\bibitem{vision_and_challenges}
S.~S. Gill \emph{et al.},
``Quantum computing: vision and challenges,''
in \emph{Quantum Computing}
(Elsevier, 2025), pp.~19--42.

\bibitem{the_quantum_internet}
A.~S. Cacciapuoti, M. Caleffi, F. Tafuri, F.~S. Cataliotti, S. Gherardini, and G. Bianchi,
``Quantum Internet: Networking Challenges in Distributed Quantum Computing,''
IEEE Network \textbf{34}, 137--143 (2020).

\bibitem{Shor_1997}
P.~W. Shor,
``Polynomial-Time Algorithms for Prime Factorization and Discrete Logarithms on a Quantum Computer,''
SIAM J. Comput. \textbf{26}, 1484--1509 (1997).

\bibitem{mohseni2024build}
M. Mohseni \emph{et al.},
``How to build a quantum supercomputer: Scaling from hundreds to millions of qubits,''
arXiv preprint arXiv:2411.10406 (2024).

\bibitem{a_survey}
M. Caleffi, M. Amoretti, D. Ferrari, J. Illiano, A. Manzalini, and A.~S. Cacciapuoti,
``Distributed quantum computing: A survey,''
Comput. Netw. \textbf{254}, 110672 (2024).

\bibitem{the_path}
R. Van Meter and S.~J. Devitt,
``The Path to Scalable Distributed Quantum Computing,''
Computer \textbf{49}, 31--42 (2016).

\bibitem{Bruzewicz_2019}
C.~D. Bruzewicz, J. Chiaverini, R. McConnell, and J.~M. Sage,
``Trapped-ion quantum computing: Progress and challenges,''
Appl. Phys. Rev. \textbf{6} (2019).

\bibitem{Klimov_2024}
P.~V. Klimov \emph{et al.},
``Optimizing quantum gates towards the scale of logical qubits,''
Nat. Commun. \textbf{15} (2024).

\bibitem{Cirac_1999}
J.~I. Cirac, A.~K. Ekert, S.~F. Huelga, and C. Macchiavello,
``Distributed quantum computation over noisy channels,''
Phys. Rev. A \textbf{59}, 4249--4254 (1999).

\bibitem{AbuGhanem2025}
M. AbuGhanem,
``Superconducting quantum computers: who is leading the future?,''
EPJ Quantum Technology \textbf{12}, 102 (2025).

\bibitem{Oreshkov2012}
O. Oreshkov, F. Costa, and \v{C}. Brukner,
``Quantum correlations with no causal order,''
Nat. Commun. \textbf{3}, 1092 (2012).

\bibitem{chiribella}
G. Chiribella, G.~M. D'Ariano, P. Perinotti, and B. Valiron,
``Quantum computations without definite causal structure,''
Phys. Rev. A \textbf{88}, 022318 (2013).

\bibitem{kraus}
S. Mondal, P. Ghosh, and U. Sen,
``Path superposition as resource for perfect quantum teleportation with separable states,''
arXiv:2505.11398 [quant-ph] (2025).

\bibitem{Rubino_2021}
G. Rubino, L.~A. Rozema, D. Ebler, H. Kristj\'{a}nsson, S. Salek, P. Allard Gu\'{e}rin, A.~A. Abbott, C. Branciard, \v{C}. Brukner, G. Chiribella, and P. Walther,
``Experimental quantum communication enhancement by superposing trajectories,''
Phys. Rev. Research \textbf{3} (2021).

\bibitem{Chiribella_2019}
G. Chiribella and H. Kristj\'{a}nsson,
``Quantum Shannon theory with superpositions of trajectories,''
Proc. R. Soc. A \textbf{475}, 20180903 (2019).

\bibitem{Gisin_2005}
N. Gisin, N. Linden, S. Massar, and S. Popescu,
``Error filtration and entanglement purification for quantum communication,''
Phys. Rev. A \textbf{72} (2005).

\bibitem{Cui_2017}
S.~X. Cui, D. Gottesman, and A. Krishna,
``Diagonal gates in the Clifford hierarchy,''
Phys. Rev. A \textbf{95} (2017).

\bibitem{Nakata_2014}
Y. Nakata and M. Murao,
``Diagonal quantum circuits: Their computational power and applications,''
Eur. Phys. J. Plus \textbf{129} (2014).

\bibitem{McKay_2017}
D.~C. McKay, C.~J. Wood, S. Sheldon, J.~M. Chow, and J.~M. Gambetta,
``Efficient $Z$ gates for quantum computing,''
Phys. Rev. A \textbf{96} (2017).

\bibitem{Vezvaee_2025}
A. Vezvaee, V. Tripathi, D. Kowsari, E. Levenson-Falk, and D.~A. Lidar,
``Virtual-$Z$ Gates and Symmetric Gate Compilation,''
PRX Quantum \textbf{6} (2025).

\bibitem{Procopio_2015}
L.~M. Procopio, A. Moqanaki, M. Ara\'{u}jo, F. Costa, I. Alonso Calafell, E.~G. Dowd, D.~R. Hamel, L.~A. Rozema, \v{C}. Brukner, and P. Walther,
``Experimental superposition of orders of quantum gates,''
Nat. Commun. \textbf{6} (2015).

\bibitem{Rubino2022experimental}
G. Rubino, L.~A. Rozema, F. Massa, M. Ara{\'{u}}jo, M. Zych, \v{C}. Brukner, and P. Walther,
``Experimental entanglement of temporal order,''
Quantum \textbf{6}, 621 (2022).

\bibitem{dof}
F.-G. Deng, B.-C. Ren, and X.-H. Li,
``Quantum hyperentanglement and its applications in quantum information processing,''
arXiv:1610.09896 [quant-ph] (2017).

\bibitem{Niemietz2021}
D. Niemietz, P. Farrera, S. Langenfeld, and G. Rempe,
``Nondestructive detection of photonic qubits,''
Nature \textbf{591}, 570--574 (2021).

\bibitem{kwiat}
P.~G. Kwiat, E. Waks, A.~G. White, I. Appelbaum, and P.~H. Eberhard,
``Ultrabright source of polarization-entangled photons,''
Phys. Rev. A \textbf{60}, R773(R)--R776(R) (1999).

\bibitem{SIMON1990165}
R. Simon and N. Mukunda,
``Minimal three-component SU(2) gadget for polarization optics,''
Phys. Lett. A \textbf{143}, 165--169 (1990).

\end{thebibliography}

\end{document}